\documentclass[12pt]{article}
\usepackage{setspace}
\usepackage{natbib}
\usepackage{graphicx}
\usepackage{lscape}
\usepackage{pdflscape}
\usepackage{afterpage}
\usepackage{pdfpages}
\usepackage{float}
\usepackage{booktabs}
\usepackage[utf8]{inputenc}
\usepackage{indentfirst}
\usepackage{arydshln}
\usepackage{arydshln}
\usepackage{tikz}
\usepackage{lipsum}
\usepackage{pgffor}
\usepackage{dsfont}
\usepackage{amsfonts}
\usepackage{amsmath}
\usepackage{xfrac}
\usepackage{amssymb}
\usepackage{graphicx}
\usepackage{url}				
\usepackage{subfig} 

\usepackage{soul}

\usepackage{graphicx}

\newtheorem{corollary}{Corollary} 

\newtheorem{proposition}{Proposition} 
\newtheorem{remark}{Remark} 

\newenvironment{proof}[1][Proof]{\noindent \textbf{#1.} }{\  \rule{0.5em}{0.5em}}

\newtheorem{assumption}{Assumption} 

\usepackage{hyperref} 
\hypersetup{
	colorlinks=true, breaklinks=true, bookmarks=true,bookmarksnumbered,
	urlcolor=red, linkcolor=red, citecolor=blue, 
	pdftitle={}, 
	pdfauthor={\textcopyright}, 
	pdfsubject={}, 
	pdfkeywords={}, 
	pdfcreator={pdfLaTeX}, 
	pdfproducer={LaTeX with hyperref and ClassicThesis} 
}

\usepackage[top=1in, bottom=1in, left=1in, right=1in]{geometry}

\usepackage{bbm}

\begin{document}

	\def\spacingset#1{\renewcommand{\baselinestretch}%
		{#1}\small\normalsize} \spacingset{1}

	
\title{ { \LARGE Randomization Inference Tests for Shift-Share Designs\footnote{We would like to thank Peter Hull for excellent comments and suggestions. }}}

\author{
Luis Alvarez\footnote{Department of Statistics, University of Sao Paulo; email: alvarez@ime.usp.br } \and Bruno Ferman\footnote{Sao Paulo School of Economics - FGV; email: bruno.ferman@fgv.br} \and Raoni Oliveira\footnote{Sao Paulo School of Economics - FGV; email: raoni.rao@gmail.com}
}

\date{}
\maketitle

	\newsavebox{\tablebox} \newlength{\tableboxwidth}
	

	\begin{center}

First draft: June 1st, 2022

\

\

\

\

\textbf{Abstract}

\end{center}

We consider the problem of inference in shift-share research designs. The choice between existing approaches that allow for unrestricted spatial correlation involves tradeoffs, varying in terms of their validity when there are relatively few or concentrated shocks, and in terms of the assumptions on the shock assignment process and treatment effects heterogeneity. We propose alternative randomization inference methods that combine the advantages of different approaches. These methods are valid in finite samples under relatively stronger assumptions, while  asymptotically valid under weaker assumptions.

\

	\noindent%

	\
	
	\noindent%
		
		\vfill

	\newpage
	\spacingset{1.45} 
	

\onehalfspacing

\section{Introduction}

Shift-share research designs consider instrumental variables that are constructed based on a common set of shocks that differentially affect different regions, depending on their exposure to those shocks. Prominent examples of papers that used this methodology include \cite{RePEc:upj:ubooks:wbsle}, \cite{RePEc:bin:bpeajo:v:23:y:1992:i:1992-1:p:1-76}, \cite{RePEc:ucp:jlabec:v:19:y:2001:i:1:p:22-64}, and \cite{Autor}.

\cite{AKM} (henceforth, AKM) show that usual standard error formulas may substantially over-state the true variability of the shift-share estimator, if regions with similar exposures to the sector-level shocks also have correlated errors. AKM and \cite{BHJ} (henceforth, BHJ) propose alternative  estimators for the asymptotic variance of the shift-share estimator that are valid under arbitrary cross-regional correlation in the regression residuals. These methods rely on an asymptotic theory in which we have a large number of sectors, and the relevance of each sector becomes asymptotically negligible. While these methods provide reliable inference in many applications, they may lead to large over-rejection when such asymptotic theory does not provide a reasonable approximation to the empirical setting \citep{Ferman_assessment}. \cite{BH} (henceforth, BH) propose another alternative,  based on the ideas of randomization inference (RI), that is valid even in finite samples. However, their approach relies on {assumptions on the shock assignment mechanism (such as, for example, knowledge of the distribution of the shocks, or that shocks are iid), which may be relatively harder to justify in some settings}. Given the advantages and disadvantages of each approach, BH state that ``\emph{the choice between RI and asymptotic approaches involves tradeoffs.}''

In this paper, we consider alternative inference methods based on RI that combine the advantages of the asymptotic methods proposed by AKM and BHJ, and of the RI method proposed by BH. The inference methods we propose  are valid in finite samples under relatively stronger assumptions, including homogeneous treatment effects, and correct specification of the distribution of the shocks up to a scale parameter. We also consider alternatives that rely on the assumptions that the distribution of shocks is symmetric around a known mean or that shocks are iid, instead of  assuming correct specification of the distribution of the shocks. At the same time, these inference methods are \emph{also} asymptotically valid when the number of sectors increases under weaker assumptions on the treatment effects heterogeneity, and even when the distribution of the shocks is misspecified.\footnote{The RI tests we consider will be asymptotically conservative whenever the conditions stated by AKM in their Appendix A.1.6 hold. Those conditions limit the correlation between treatment effect heterogeneity and exposure weights.}  Therefore, we provide inference methods for shift-share designs that are valid under relatively  stronger assumptions in finite samples, but that we can relax those assumptions once the number of sectors increases. This eliminates the tradeoffs between RI and asymptotic approaches mentioned by BH.

Our approaches build on a large literature that studies the use of RI methods in other settings, and considers RI with studentized test statistics that are valid under stronger assumptions (or, alternatively, for inference on sharper null hypotheses) in finite samples, and also asymptotically valid under weaker assumptions (or, alternatively, for inference on less stringent null hypotheses). See, for example, \cite{Janssen1997}, Chapter 15 of \cite{Lehmann2005}, \cite{Chung2013}, \cite{Bugni2018}, \cite{Wu2020}, \cite{Ferman_matching}.

\section{Main results}

\subsection{Setting}

	For an outcome of interest $Y$, we consider the structural model
	
	\begin{equation}
		Y(x,\epsilon) = \beta \cdot x + \epsilon,
	\end{equation}
	where $x$ denotes a treatment of interest, and $\epsilon$ are the remaining determinants of $Y$. We consider for simplicity the case without a constant and without other covariates. However, all our results remain valid for a more general setting. For a sample of $N$ units, observed outcomes are given by
	
	\begin{equation}
		Y_i = Y(X_i, \epsilon_i) \quad i=1,\ldots,N,
	\end{equation}
	where $\{(X_i, \epsilon_i)\}_{i=1}^N$ are random variables. We have access to a shift-share instrument constructed as
	
	\begin{equation}
		Z_i = \boldsymbol{s}_i' g,
	\end{equation}
where $\boldsymbol{s}_i \in \mathbb{R}^J_+$ are a set of exposures of $i$ and $g \in \mathbb{R}^J$ are a set of sector shocks. Let $\boldsymbol{S} = [\boldsymbol{s}_1,\boldsymbol{s}_2,\ldots \boldsymbol{s}_N]'$ and $\boldsymbol{\epsilon} = (\epsilon_1,\ldots, \epsilon_N)'$.  We consider the following assumption on the distribution of the shocks, which is standard in the shift-share design literature (see AKM and BHJ).

\begin{assumption}[Shock exogeneity]  \label{Assumption_shock}
$\mathbb{E}[g|\boldsymbol{S},\boldsymbol{\epsilon}] = 0$.
\end{assumption}

Assumption \ref{Assumption_shock} imposes that shocks are mean-independent from unobserved determinants, conditional on exposures, with a common mean. More generally, we could have assumed that there exists $\mu \in \mathbb{R}$ such that $\mathbb{E}[g_j|\boldsymbol{S},\boldsymbol{\epsilon}] = \mu$ for every $j=1,\ldots, J$. In this case, the researcher may conduct inference by working with statistics that depend on \emph{demeaned shocks} $\tilde{g}_j = g_j -\bar{g}$ (e.g. the shift-share estimator constructed with demeaned shocks). This is the solution proposed by BH to inference in linear shift-share designs where the sum of exposures may be uneven across units.\footnote{Our results remain valid in such setting. Specifically, valid finite sample inference under correct specification of the shock assignment mechanism (Proposition \ref{Prop_finite}) would solely require that shocks are correctly specified up to a common location shift. Another alternative would be to control for the sum of the exposures, as proposed by BHJ.}

In our setting, the shift-share estimator is given by

\begin{equation}
	\hat{\beta}_{SS} = \hat{\beta}_{\text{IV}}( \boldsymbol{X}, \boldsymbol{Y},\boldsymbol{Z}) = \frac{\sum_{i=1}^N Z_i Y_i}{\sum_{i=1}^N Z_i X_i} = \beta + \frac{\sum_{i=1}^N (\boldsymbol{s}_i'g)\epsilon_i}{\sum_{i=1}^N (\boldsymbol{s}_i'g) X_i},
\end{equation}
where $\boldsymbol{X} = (X_1,\ldots, X_N)'$, $\boldsymbol{Y} = (Y_1,\ldots, Y_N)'$ and $\boldsymbol{Z} = (Z_1,\ldots, Z_N)'$. All our results remain valid if we consider the reduced-form case, in which case we set $X_i = Z_i$. Also, some of the results we present in Section \ref{Sec_RI} are only valid for the reduced-form case.

AKM and BHJ show that, when shocks are independent and the importance of each sector becomes asymptotically negligible, then, as $J, N \to \infty$, 

\begin{equation}
	\mathbb{P}[V_{SS}^{-1/2}(\hat{\beta}_\text{SS} - \beta)\leq c|\boldsymbol{S},\boldsymbol{\epsilon}]  \overset{p}{\to} \Phi(c),
\end{equation} 
for every $c \in \mathbb{R}$, where $V_{SS}=\frac{\left(\sum_{i=1}^N \epsilon_i\boldsymbol{s}_i\right)' \mathbb{V}[g|\boldsymbol{S},\boldsymbol{\epsilon}]\left(\sum_{i=1}^N \epsilon_i\boldsymbol{s}_i\right)}{(\sum_{i=1}^N Z_i X_i)^2}$. This representation motivates the variance estimators proposed by AKM and BHJ. While their approaches provide reliable inference in empirical applications with many sectors, we may have relevant size distortions when there are few or concentrated sectors.

\subsection{Randomization inference in shift-share designs } \label{Sec_RI}

Given that the inference methods proposed by AKM and BHJ may not work well in some applications when there are few or concentrated sectors, we consider the use of RI in this setting. BH propose randomization-based inference in shock-based designs for a more general setting in which we have non-random exposure to exogenous shocks, where shift-share designs would be a particular example. Differently from BH,  by focusing on shift-share design applications we are able to consider RI tests that are valid under relatively stronger assumptions in finite samples (similar to the approach proposed by BH), but that are also valid under weaker assumptions when the number of sectors increases. The main  reason is that, in the setting we consider, the shift-share estimator has well-stablished asymptotic results (AKM, BHJ), so we are able to consider a studentized test statistic. In contrast, many of the settings considered by BH do not have well-stablished asymptotic results.

\subsubsection{Finite-sample results}

Suppose our goal is to test the null that $\beta = b$ against either a unilateral or bilateral alternative. Let $\hat{T} = \bar{T}(g, \boldsymbol{S}, \boldsymbol{X}, \boldsymbol{Y})$ be a test statistic, where large values of $\hat{T}$ constitute evidence against the null. We assume the following condition on the test statistic. 

\begin{assumption}
\label{Ass_stat}
{Under the null}, the map $\bar{T}$ satisfies $\bar{T}(g, \boldsymbol{S},  \boldsymbol{Y},\boldsymbol{X}) = T(g, \boldsymbol{S}, \boldsymbol{Y} - \boldsymbol{X}b)$ for some other map $T$.

\end{assumption} 

That is, under the null, the test-statistic depends on  $\boldsymbol{Y}$ and $\boldsymbol{X}$ solely through the ``null-imposed residuals'' $\boldsymbol{e}_b = \boldsymbol{Y} - \boldsymbol{X}b$.

Suppose now that the researcher has a guess on the shock assignment mechanism, i.e on the conditional probabilities $\mathbb{P}[g  \leq v|\boldsymbol{S},\boldsymbol{\epsilon}]$ for each $v \in \mathbb{R}^J$. Let us denote such guess by a conditional distribution function $\mathbb{H}(\cdot|\boldsymbol{s},\boldsymbol{e})$ which specifies, for each $(\boldsymbol{s},\boldsymbol{e})$ in the support of $(\boldsymbol{S},\boldsymbol{\epsilon})$, a cummulative distribution function $\mathbb{H}(\cdot|\boldsymbol{s},\boldsymbol{e})$ on $\mathbb{R}^J$. In this case, the researcher is able to compute critical values by analysing the quantiles of

\begin{equation}
	H_b(c|\boldsymbol{S},\boldsymbol{X},\boldsymbol{Y}) = \int \mathbf{1}_{\{T(g, \boldsymbol{S}, \boldsymbol{e}_b) \leq c\}} \mathbb{H}[d g|\boldsymbol{S},\boldsymbol{e}_b].
\end{equation}

Such quantity is easily estimable by simulation. Indeed, if we are able to draw $L$ independent draws $g^*_l$, $l=1,\ldots, L$, from $\mathbb{H}[\cdot |\boldsymbol{S},\boldsymbol{e}_b]$, then $H_b(c|\boldsymbol{S},\boldsymbol{X},\boldsymbol{Y})$ may be estimated as
\begin{equation}
	\hat{H}_b(c|\boldsymbol{S},\boldsymbol{X},\boldsymbol{Y}) = \frac{1}{L+1}\left(1+\sum_{l=1}^L \mathbf{1}_{\{T(g^*_l, \boldsymbol{S}, \boldsymbol{e}_b) \leq c\}}\right).
\end{equation}

Clearly, if the shock-assignment process is correctly specified, then the procedure above provides valid inference.

\begin{proposition} \label{Prop_finite}

Suppose that Assumption \ref{Ass_stat} holds, and that $\mathbb{P}[g \leq \cdot|\boldsymbol{S},\boldsymbol{\epsilon}] = \mathbb{H}[\cdot|\boldsymbol{S},\boldsymbol{\epsilon}]$. Then, under the null $\beta = b$,
	
	$$	H_b(c|\boldsymbol{S},\boldsymbol{X},\boldsymbol{Y}) = \mathbb{P}[\hat{T}\leq c|\boldsymbol{S},\boldsymbol{\epsilon}].$$

Consequently, a test that rejects the null if $\hat{T}$ exceeds the $1-\alpha$ quantile of $H_b(\cdot|\boldsymbol{S},\boldsymbol{X},\boldsymbol{Y})$ is (conditional on $(\boldsymbol{S},\boldsymbol{X},\boldsymbol{Y})$) level $\alpha$, where $\alpha \in (0,1)$. Similarly, a test that rejects the null if $\hat{T}$ exceeds the $1 +\frac{2}{L+1} -\frac{\lceil\alpha (L+1)\rceil}{L+1}$ quantile of $\hat{H}_b$ is conditionally level $\alpha$.

\end{proposition}

The result presented above is similar to Proposition S.3 from BH, with a  couple of minor differences. First, we allow the shock assignment mechanism to depend on the non-observables. This way, we allow for the distribution of $S$ to depend on $\boldsymbol{\epsilon}$ (though in practice we expect that applied researchers would  rarely choose a $\mathbb{H}[\cdot|\boldsymbol{S},\boldsymbol{e}_b]$  that depends on $\boldsymbol{e}_b$). Also, we provide level guarantees with a finite number of simulations. We present details of the proof in Appendix \ref{Proof_Prop_finite}.

In their paper, BH consider basing inference on the test statistic

\begin{equation}
{\hat T_0 = } \bar{T}_0(g, \boldsymbol{S}, \boldsymbol{X}, \boldsymbol{Y}) = \frac{1}{N} \sum_{i=1}^N (g'\boldsymbol{s}_i) (Y_i - b X_i),
\end{equation}
which depends on $(\boldsymbol{X},\boldsymbol{Y})$ solely through $\boldsymbol{e}_b$, so Assumption \ref{Ass_stat} holds.

In contrast, we consider the test-statistic

\begin{equation}
\hat{T}_1 =	\bar{T}_1(g, \boldsymbol{S}, \boldsymbol{X}, \boldsymbol{Y}) = \frac{\hat{\beta}_{SS}- b}{\sqrt{\hat{V}_{N,b}}},
\end{equation}
where $\hat{V}_{N,b}$ is the \textbf{null-imposed} variance estimator from {AKM and BHJ,\footnote{In a setting without an intercept or controls, BHJ show their variance estimator collapses to AKM's.}}

$$\hat{V}_{N,b} = \frac{\sum_{j=1}^J\left(\sum_{i=1}^N (Y_i - bX_i)\boldsymbol{s}_{ij}\right)^2 g_j^2}{(\sum_{i=1}^N Z_i X_i)^2}.$$

Observe that, under the null, the above map satisfies
\begin{equation}
	\bar{T}_1(g, \boldsymbol{S}, \boldsymbol{X}, \boldsymbol{Y}) = \frac{\sum_{i=1}^N (g'\boldsymbol{s}_i)(Y_i - X_i b)}{\sqrt{\sum_{j=1}^J\left(\sum_{i=1}^N (Y_i - bX_i)\boldsymbol{s}_{ij}\right)^2 g_j^2}} \equiv T_1(g, \boldsymbol{S}, \boldsymbol{e}_b),
\end{equation}
so this test statistic also satisfies Assumption \ref{Ass_stat}. This corresponds to a rescaled version of the test statistic in BH.

 Inference may be conducted as follows:

\paragraph{Algorithm 1}
\begin{enumerate}

	\item For simulations $l=1,\ldots,L$:
	\begin{enumerate}
		\item Draw $g^*_l \sim \mathbb{H}[\cdot|\boldsymbol{S},\boldsymbol{e}_b]$.
		\item Construct simulated instruments, $Z^*_{il} = \boldsymbol{s}_i'g^*_l$.
		\item Run the shift-share IV estimator and null-imposed standard errors using the original data $Y_i$ and $X_i$ with artificial instruments $Z_{il}^*$. Construct the $t$-test based on the obtained values.
	\end{enumerate}
	\item Reject the null if the observed test statistic is at the tails of the simulated distribution.
\end{enumerate}

We adopt null-imposed standard errors in the test statistic $\hat T_1$, because otherwise it would depend on the endogenous regressor $X_i$, and we do not model assignment of these. If, however, $X_i = Z_i$, such problem disappears, and we may consider the test statistic
\begin{equation}
{\hat T_2 =  }	 \bar{T}_2(g, \boldsymbol{S}, \boldsymbol{Y}) = \frac{\hat{\beta}_{SS} - b}{\sqrt{\hat{V}_{F}}},
\end{equation}
where $\hat{V}_{F} =  \frac{\sum_{j=1}^J\left(\sum_{i=1}^N (Y_i - \hat{\beta}_{SS}X_i)\boldsymbol{s}_{ij}\right)^2 g_j^2}{(\sum_{i=1}^N X_i^2)^2}$ is the AKM or BHJ standard errors without imposing the null. Under the null, such statistic may be written as

\begin{eqnarray} \nonumber
	\bar{T}_2(g, \boldsymbol{S}, \boldsymbol{Y}) &=& \frac{\sum_{i=1}^N(g'\boldsymbol{s}_i)'\boldsymbol{e}_b}{\sqrt{{\sum_{j=1}^J\left(\sum_{i=1}^N ((\boldsymbol{s}_i'g)[b-\hat{\beta}_{IV}(\boldsymbol{S}g, b\boldsymbol{S}g + \boldsymbol{e}_b, \boldsymbol{S}g)] + \boldsymbol{e}_{i,b})\boldsymbol{s}_{ij}\right)^2 g_j^2}}} \\
	&\equiv& T_2(g, \boldsymbol{S}, \boldsymbol{e}_b).
\end{eqnarray}

Therefore, when we are in the case in which $X_i = Z_i$, we have that this test statistic satisfies Assumption \ref{Ass_stat}. Note, however, that this assumption would not be satisfied for this test statistic if we considered the case in which $X_i \neq Z_i$.

In this case,  inference may be conducted as follows:

\paragraph{Algorithm 2}
\begin{enumerate}
	\item For a given $b$, compute $\boldsymbol{e}_b = \mathbf{Y} -  \mathbf{X} b$.
	\item For simulations $l=1,\ldots,L$:
	\begin{enumerate}
		\item Draw $g^*_l \sim \mathbb{H}[\cdot|\boldsymbol{S},\boldsymbol{e}_b]$.
		\item Construct data $\boldsymbol{Y}_l^* = b \boldsymbol{S} g^*_l + \boldsymbol{e}_b$.
		\item Run the shift-share regression and shock-robust standard errors using the artificial data $\boldsymbol{Y}_l^*$, $g_l^*$ and $\boldsymbol{S}$. Construct the $t$-test based on the obtained values.
	\end{enumerate}
	\item Reject the null if the observed test statistic is at the tails of the simulated distribution.
\end{enumerate}

Following Proposition \ref{Prop_finite}, this inference procedure would be valid for settings in which $X_i = Z_i$.

\begin{remark}[Scale-invariance of $\bar{T}_1$ and $\bar{T}_2$]
\normalfont
	We observe that, when inference is based on the test-statistics  $\hat{T}_1$ or $\hat{T}_2$, the requirement in Proposition \ref{Prop_finite} may be weakened to: the distribution of shocks $g$ is correctly specified, up to multiplication of $g$ by a positive scalar. Indeed, test statistics $\hat{T}_1$ and $\hat{T}_2$ are invariant to multiplication of the shocks by a common positive constant. This contrasts with the test statistic $\hat{T}_0$, which requires the researcher to correctly specify the scale of shocks.
\end{remark}

\begin{remark}[Group transformations]
\normalfont
	Instead of assuming that the shock-assignment mechanism is known, an alternative would be to consider a group $\boldsymbol{H}$ of transformations on $\mathbb{R}^J$ such that, under the null, for any $h \in \boldsymbol{H}$, $\bar{T}(h(g), \boldsymbol{S}, \boldsymbol{\epsilon})|\boldsymbol{S},\boldsymbol{\epsilon} \overset{d}{=} \bar{T}(g, \boldsymbol{S}, \boldsymbol{\epsilon})|\boldsymbol{S},\boldsymbol{\epsilon}$. In these settings, it follows from well-established results on randomization tests \citep[Theorem 15.21]{Lehmann2005} that the procedure described in Proposition \ref{Prop_finite} remains valid if simulated  shocks are constructed as $g^* = \boldsymbol{h}(g)$, where $\boldsymbol{h} \sim \operatorname{Uniform}(\boldsymbol{H})$, independently from the data. For example, if, conditional on $(\boldsymbol{S}, \boldsymbol{\epsilon})$, shocks were assumed independently drawn from symmetric distributions with known common symmetry point $m$, then one could take the group of transformations to be recentred sign changes, i.e. $h(g) = \kappa \odot (g - m \cdot \iota_J) + m \cdot \iota_J$ for $\kappa \in \{-1,1\}^J$, where $\odot$ denotes entry-by-entry multiplication and $\iota_J$ is a $J$ dimensional vector of ones.\footnote{ If the symmetry point were estimated (for example, by using the sample mean as an estimator of $m$), then the simulation procedure would no longer retain finite sample validity. In this case, conservative inference could be conducted by computing p-values under different choices of $m$, as $m$ varies over a valid confidence set, and then taking the supremum and adding one minus the confidence of the confidence set to it \citep{Berger1994}. See Proposition S6 in BH for details.} {BH consider this kind of simulations in their Appendix D4.} Similarly, if, conditional on $(\boldsymbol{S}, \boldsymbol{\epsilon})$, shocks were assumed to be iid, then one could take the group to be the set of permutations of a $J$-dimensional vector,  as also discussed by BH.
\end{remark}

\subsubsection{Asymptotic results}

In this section, we consider the asymptotic properties of the simulation-based approach. We consider the properties of the tests in a framework where the number of sectors, $J$, is large. The number of units, $N$, is (implicitly) indexed by $J$, and is also allowed to grow. 

We adopt a finite population perspective and allow for treatment effects to vary by unit. Formally, for a given $J \in \mathbb{N}$, potential outcomes are given by

\begin{equation}
	Y_{i, J}(x, \epsilon) = \beta_{i,J} x + \epsilon, \quad i=1,\ldots, N,
\end{equation}
and observed outcomes are given by
\begin{equation}
	Y_{i,J} = Y_{i,J}(X_{i,j}, \epsilon_{i,J}) = \beta_{i,J} X_{i,J} + \epsilon_{i,J},
\end{equation}
whereas the instrument is given by $Z_{i,J} = \boldsymbol{s}_{i,J}'g_{J}$. We will treat the $\epsilon_{i,J}$, $\beta_{i,J}$ and $\boldsymbol{s}_{i,J}$ as nonrandom throughout -- the only source of randomness stems from the assignment of $g_J$ and the treatment $X_{i,J}$. In other words, we follow a ``design-based'' approach. In this setting, the shift-share identification assumption is written as follows.

\begin{assumption}[Shock-exogeneity] \label{Ass_exog_large}
		$\mathbb{E}[g_J] = 0$.
\end{assumption}

We rewrite the outcome model as

\begin{equation}
	Y_{i,J} = \beta_J X_{i,J} + \eta_{i,J} + \epsilon_{i,j},
\end{equation}
where

\begin{equation}
	\beta_J = \frac{\sum_{i=1}^N \mathbb{E}[X_{i,J} Z_{i,J}] \beta_{i,J}}{\sum_{i=1}^N \mathbb{E}[X_{i,J} Z_{i,J}]},
\end{equation}
and 
\begin{equation}
	\eta_{i,J} = (\beta_{i,J} - \beta_J)X_{i,J}.
\end{equation}

We consider the goal of the researcher to be to conduct inference on $\beta_J$, an affine combination of individual treatment effects. Specifically, she would like to test the null that $\beta_J = b_J$, and for that she uses one of the procedures described in the previous section.

Following AKM and BHJ, we put $v_J = \sum_{j=1}^J|\sum_{i=1}^N {s}_{i,j,J}|^2$ and assume $v_J \to \infty$. In the next proposition, we provide conditions for (conditional) asymptotic normality of $T^*_1 = T_1(g^*_J, \boldsymbol{S}, \boldsymbol{Y} - b_J \boldsymbol{X})$, the test statistic ${T}_1$ constructed under simulated shocks $g^*_J$.

\begin{proposition}[Asymptotic normality of $T_1^*$ statistic] \label{Prop_asympt}
	Assume that, conditional on $g_J$ and $\boldsymbol{X}$, simulated shocks $g^*_{j,J}$ are drawn independently across $j$ from distributions (not necessarily identical) satisfying:

	\begin{enumerate}
		\item[(i)]$\frac{1}{\sqrt{v_J}}\sum_{j=1}^J \left( \sum_{i=1}^N [\epsilon_{i,J}+\eta_{i,J} + (\beta_J-b_J)X_{i,J}]\boldsymbol{s}_{i,j,J}\right)\mathbb{E}[g_{j,J}^*|g_J,\boldsymbol{X}]=o_p(1)$;
		\item[(ii)] $\sum_{j=1}^J\frac{1}{v_J}\left( \sum_{i=1}^N [\epsilon_{i,J}+\eta_{i,J} + (\beta_J-b_J)X_{i,J}]\boldsymbol{s}_{i,j,J}\right)^2\mathbb{E}[g_{j,J}^{*2}|g_J,\boldsymbol{X}] \overset{p}{\to} \sigma^2_* $, where $\sigma^2_* > 0$; and
		\item[(iii)] $\sum_{j=1}^J\frac{1}{v_J^2}\left( \sum_{i=1}^N [\epsilon_{i,J}+\eta_{i,J} + (\beta_J-b_J)X_{i,J}]\boldsymbol{s}_{i,j,J}\right)^4\mathbb{E}[g_{j,J}^{*4}|g_J,\boldsymbol{X}] \overset{p}{\to} 0 $.
	\end{enumerate} 
	
	Then, the simulated distribution converges in distribution to a standard normal, in probability, i.e. $
			\mathbb{P}[T_1^* \leq c|\boldsymbol{X},g_J]\overset{p}{\to} \Phi(c)$, for every $c \in \mathbb{R}$.

\end{proposition}

We present details of the proof in Appendix \ref{Proof_Prop_asympt}. Proposition \ref{Prop_asympt} provides high-level conditions for conditional asymptotic normality of the simulated statistic.  In Appendix \ref{Ver_prop_asympt}, we show that these conditions are satisfied for three examples of simulation distributions: (i) when we sample with replacement from the (recentered) empirical distribution of shocks; (ii) when we consider shocks iid $N(0,1)$, independently from $\mathbf{X}$ and $g_J$, and; (iii) when we consider sign-changes of observed shocks.\footnote{{In the Appendix, we consider sign changes without recentering shocks ($m=0$). We note, however, that convergence would hold for any choice of recentering parameter $m$, including the case in which it is misspecified, and the case in which $m$ is replaced by an estimator such as the  sample mean of shocks;  {provided we work with the shift-share estimator that uses demeaned shocks (as per footnote 2)}.}} The crucial point is that we consider a studentized test statistic, so  the simulated test statistic is asymptotically $N(0,1)$. In contrast, if we considered alternative test statistics, such as $\hat T_0$, then we would not reach this conclusion. Studentizing the test statistic using robust standard errors would also generally not work.

As a byproduct of Proposition \ref{Prop_asympt}, whenever inference based on $\hat{T}_1$ and normal critical values provides asymptotically conservative inference, the simulation-based approach will also lead to asymptotically conservative inference. We summarize this fact in the corollary below.

\begin{corollary} \label{Cor}
	Suppose that, under the null $ \beta_J = b_J$, there exists $ v\geq 1$ such that, for every $c \in \mathbb{R}$, $\mathbb{P}[\hat{T}_1 \leq c] \to \Phi(v\cdot c)$. Assume that the conditions in Proposition \ref{Prop_asympt} hold under the null. Then the simulation-based approach to inference of Algorithm 1 will be asymptotically conservative, in the sense that, under the null, the probability of rejecting the null converges to a number smaller than the nominal significance level.
\end{corollary}

When there is no treatment effect heterogeneity, it follows that, under the conditions in AKM and BHJ, $v=1$. These conditions include that shocks are independent, that the number of sectors increase, and that the relevance of sectors are asymptotically negligible (AKM and BHJ consider alternatives that relax the assumption that shocks are independent, and we discuss that in Remark \ref{Rem_cluster}). In this case, inference based on the simulation approach is asymptotically size $\alpha$. More generally, when there is treatment effect heterogeneity, AKM provide sufficient conditions for inference based on $\hat{T}_1$ and normal critical-values being conservative (i.e. $v \geq 1$). These conditions limit the correlation between treatment effect heterogeneity and exposure weights. In this case, our simulation-based approach will also lead to conservative inference. Notice that, in contrast to our finite sample results, which require homogeneous treatment effects, asymptotically our method may be able to provide conservative inference under treatment effect heterogeneity.

\begin{remark}
\normalfont \label{Rem_power}
	We note that the statement of Proposition \ref{Prop_asympt} does not require the null to be true. Specifically, if the conditions in Proposition \ref{Prop_asympt} can be shown to be valid under a given sequence of alternatives,\footnote{See Appendix \ref{Ver_prop_asympt} for sufficient conditions in our three examples of simulation distributions.} then it follows that the distribution of the simulated statistic converges to a standard normal along such sequence. In this case, the power of the null-imposed t-test and our simulation-based approach coincide asymptotically along this sequence.
\end{remark}

\begin{remark}
\normalfont
\label{Rem_cluster}

Suppose that instead of assuming that shocks are independent, we consider that we have clusters of shocks that are independent, but that there may be correlation between shocks within the same cluster. In this case,  our results from Proposition \ref{Prop_asympt} and Corollary \ref{Cor} should remain valid if we studentized the test statistic using AKM and BHJ standard errors  with clusters of shocks (with the null imposed), provided the number of clusters is large. We may also consider using a distribution for the simulated shocks that allows for correlation within clusters.

\end{remark}

Next, we analyze the test statistic $\hat T_2$. In this case, since the shift-share estimator is being recomputed across samples and then used in the calculation of the standard error, we need to ensure that $\hat{\beta}_{SS}^*$, the simulated shift-share estimator from Algorithm 2, is consistent at a given rate. In addition to the assumptions in Proposition \ref{Prop_asympt}, we require a ``strong simulated shock'' assumption that ensures that the variance of the simulated shift-share regressor does not vanish asymptotically; as well as conditions that ensure the estimation error of the standard error vanishes. We state these requirements in the proposition below:

\begin{proposition} \label{Prop_asympt2}
	Suppose, in addition to the assumptions in Proposition \ref{Prop_asympt}, that: (i) $\frac{1}{N} \sum_{i=1}^N (\boldsymbol{s}_{i,J}'g_J^*)^2 \overset{p}{\to} \pi^* > 0$. Then  $\frac{N}{\sqrt{v_J}}(\hat{\beta}^*_{SS} -b_J) = O_P(1)$. Moreover, if we assume that:
	\begin{enumerate}
		\item[(ii)] 	\begin{equation*}
			\begin{aligned}
					\frac{1}{v_J}	\sum_{j=1}^J \Bigg(\sum_{i=1}^N \sum_{l=1}^N s_{i,j,J}s_{l,j,J}[(\boldsymbol{s}_l'g^*_{J})(\epsilon_{i,J}+\eta_{i,J} + (\beta_J-b_J)X_{i,J}) + \\ (\boldsymbol{s}_i'g^*_{J})(\epsilon_{l,J}+\eta_{l,J}+  (\beta_J-b_J)X_{i,J})]\Bigg) g_{j,J}^{*2} =  o_p\left(\frac{N}{\sqrt{v_J}}\right); \text{and}
			\end{aligned}
		\end{equation*}
		\item[(iii)] 	\begin{equation*}
			\frac{1}{v_J}	\sum_{j=1}^J \Bigg(\sum_{i=1}^N \sum_{l=1}^N [ (\boldsymbol{s}_i'g^*_{J}) (\boldsymbol{s}_l'g^*_{J})s_{i,j,J}s_{l,j,J} ]\Bigg) g_{j,J}^{*2} = o_p\left(\frac{N^2}{{v}_J}\right),
		\end{equation*}
	\end{enumerate}
	we may then conclude that $\mathbb{P}[T_2^* \leq c|\boldsymbol{X},g_J] \overset{p}{\to} \Phi(c)$ for every $c \in \mathbb{R}$.

\end{proposition}

We present details of the proof in Appendix \ref{Proof_Prop_asympt2}. In Appendix \ref{Ver_prop_asympt2}, we discuss assumptions (i)-(iii) of the proposition in the context of our three examples of simulation distributions.

\begin{corollary} 
\label{Cor2}
	Suppose that, under the null $\beta_J=b_J$, there exists $ v\geq 1$ such that, for every $c \in \mathbb{R}$, $\mathbb{P}[\hat{T}_2 \leq c] \to \Phi(v\cdot c)$. Assume that the conditions in Proposition \ref{Prop_asympt2} hold under the null. Then the simulation-based approach to inference provided by Algorithm 2 will be asymptotically conservative, in the sense that, under the null, the probability of rejecting the null converges to a number smaller than the nominal significance level.
\end{corollary}

\begin{remark}
\normalfont

Remarks \ref{Rem_power} and \ref{Rem_cluster}  also apply to Proposition \ref{Prop_asympt2} and Corollary \ref{Cor2} when we consider the test statistic $\hat T_2$.

\end{remark}

\section{Conclusions}

We consider the problem of inference in shift-share research designs. There are two main existing approaches that allow for unrestricted spatial correlation. The RI approach is valid even with relatively few or concentrated shocks, but relies on relatively strong assumptions on the shock assignment process and on treatment effect heterogeneity. In contrast, the asymptotic approach relies on weaker assumptions on the shock assignment process and on treatment effect heterogeneity, but asymptotic approximations may be inaccurate in some applications.

We propose alternative RI methods that combine the advantages of both approaches. More specifically, the inference methods we propose are exact under relatively strong assumptions, and also asymptotically valid under weaker assumptions. The latter is achieved through studentization, which ensures convergence of the simulated distribution of the test-statistic to a standard normal under mild regularity conditions.



\singlespace

\renewcommand{\refname}{References} 

\bibliographystyle{apalike}
\bibliography{bib/bib.bib}

\pagebreak

\appendix

\setcounter{table}{0}
\renewcommand\thetable{A.\arabic{table}}

\setcounter{figure}{0}
\renewcommand\thefigure{A.\arabic{figure}}

\section{Online Appendix}

\subsection{Proof of main results}

\subsubsection{Proof of Proposition \ref{Prop_finite}}
\label{Proof_Prop_finite}
\begin{proof}
	The first assertion is immediate. We prove the second assertion. Recall that the quantile function of a distribution function $F$ is given by $Q_F(u) = \inf \{z \in \mathbb{R}: F(z) \geq u\}$. Consequently, under the null, $\mathbb{P}[\hat{T} >Q_{H_b}(1-\alpha)|\boldsymbol{S},\boldsymbol{X},\boldsymbol{Y}] = 1 - H_b(Q_{H_b}(1-\alpha)|\boldsymbol{S},\boldsymbol{X},\boldsymbol{Y})\leq \alpha$, as desired. Similarly, by the definition of quantile function, $\hat{T} > Q_{\hat{H}_b}\left(1 + \frac{2}{L+1}- \frac{\lceil\alpha (L+1)\rceil}{L+1}\right)$ implies that $\hat{H}_b\left(\hat{T}|\boldsymbol{S},\boldsymbol{X},\boldsymbol{Y}\right) \geq 1 +\frac{2}{L+1} - \frac{\lceil\alpha (L+1)\rceil}{L+1}  $. Observe that $\hat{H}_b\left(\hat{T}|\boldsymbol{S},\boldsymbol{X},\boldsymbol{Y}\right) \geq  1 +\frac{2}{L+1} - \frac{\lceil\alpha (L+1)\rceil}{L+1} $ implies the sample statistic is strictly greater than $ (L+1) - \lceil\alpha(L+1)\rceil$ simulated draws. Under the null, the latter event occurs with probability $\mathbb{E}[(1 -H_b(T^*_{[(L+1) - \lceil\alpha (L+1)\rceil]}|\boldsymbol{S}, \boldsymbol{X}, \boldsymbol{Y}))|\boldsymbol{S}, \boldsymbol{X}, \boldsymbol{Y}]$, where $T^*_{[k]}$ is the $k$-th order statistic associated with the $L$ simulations. Using the quantile representation of a random variable and the definition of a quantile function, one obtains that:
	
	$$\mathbb{E}[H_b(T^*_{[(L+1)- \alpha (L+1)]}|\boldsymbol{S}, \boldsymbol{X}, \boldsymbol{Y})|\boldsymbol{S}, \boldsymbol{X}, \boldsymbol{Y}] \leq \mathbb{E}[U_{[(L+1)- \lceil\alpha(L+1)\rceil]}] \leq \frac{(L+1) - \alpha(L+1)}{L+1} = 1 - \alpha,$$
	where $U_{[k]}$ is the $k$-th order statistic of a sample of $L$ independent normals, which has expected value $k/(L+1)$.
\end{proof}

\subsubsection{Proof of Proposition \ref{Prop_asympt}}
\label{Proof_Prop_asympt}

	\begin{proof}
	
	Observe that, under the null $T^*_1$ is written as
	
	\begin{equation}
		\frac{\sum_{i=1}^N (\boldsymbol{s}_{i,J}'g^*_J) (\epsilon_{i,J}+ \eta_{i,J} + (\beta_J-b_J)X_{i,J})}{\sqrt{\sum_{j=1}^J \left(\sum_{i=1}^N [\epsilon_{i,J}+\eta_{i,J}]\boldsymbol{s}_{i,j,J}\right)^2 g_{j,J}^{*2}} }.
	\end{equation}
	Let $\mathbb{E}_*$ denote the conditional expectation on $g_J$ and $\boldsymbol{X}$. We begin by showing the squared denominator, rescaled by $v_L$, which we denote by $D_J^2/v_J$ is consistent. To see this, we note that, by Assumption (iii) in the statement of the Proposition:
	\begin{equation}
		\begin{aligned}
		\mathbb{E}_*|D_J^2 - \mathbb{E}_* D_J^2|^2\leq \mathbb{E}_*\left|\sum_{j=1}^J \left(\sum_{i=1}^N [\epsilon_{i,J}+\eta_{i,J} + (\beta_J-b_J)X_{i,J}]\boldsymbol{s}_{i,j,J}\right)^2 [g_{j,J}^{*2} - \mathbb{E}_*g_{j,J}^2 ]\right|^2 \leq \\ \leq   4 \sum_{j=1}^J\left(\sum_{i=1}^N [\epsilon_{i,J}+\eta_{i,J}+(\beta_J-b_J)X_{i,J}]\boldsymbol{s}_{i,j,J}\right)^4 \mathbb{E}_*|g^{*4}_{j,J}| = o_p(v^2_J),
		\end{aligned}
	\end{equation}
which proves, by application of the conditional Markov inequality and bounded convergence, that $\frac{1}{v_J}|D_J^2 - \mathbb{E}_*[D_J^2]| = o_p(1)$. Combined with Assumption (ii) in the statement of the Proposition, we obtain that $D_J^2/v_J  \overset{p}{\to} \sigma^2_*$.

Next, we need to show that the distibution of the numerator, rescaled by  $\sqrt{v_J}$, which we denote by $N_J/\sqrt{v_J}$, converges to a normal distibution. We first note that, passing through a subsequence if needed, the convergence in probability requirements in the statement of the proposition, as well as the consistency of the denominator previously shown, may be taken as almost-sure convergence.\footnote{By the fact that a sequence converges in probability if, and only if, every subsequence admits a further subsequence that converges almost surely \citep[Theorem 20.10]{Billingsley1995}} We are thus able to apply a CLT for triangular arrays \citep[Theorem 3.4.10]{Durrett2019}. Indeed, we observe that

\begin{equation}
	N_J/\sqrt{v_J} = \sum_{j=1}^J \omega_{j,J}g_{j,J}^*, 
\end{equation}
with $\omega_{j,J} = \frac{1}{\sqrt{v_J}}(\sum_{i=1}^N [\epsilon_{i,J}+\eta_{i,J} + (\beta_J-b_J)X_{i,J}]s_{i,j,J})$. Notice that, $\sum_{j=1}^J \omega_{j,J}\mathbb{E}_*[g_{j,J}^*] \to 0$ a.s. and $\sum_{j=1}^J \mathbb{E}_*[\omega_{j,j}^2 g_{j,J}^2] \to \sigma^2_*$ a.s.. It thus suffices to verify the Lindeberg condition in our problem. Fix $\epsilon > 0$. We have:

\begin{equation}
	\sum_{j=1}^J \mathbb{E}_*[\omega_{j,J}^2 g^{*2}_{j,J} \mathbf{1}_{|\omega_{j,J} g_{j,J}| > \epsilon}] \leq 
	\frac{1}{\epsilon^2}\sum_{j=1}^J \omega_{j,J}^4 \mathbb{E}_*[g_{j,J}^{*4}]   \to 0,
\end{equation}
where the (a.s.) convergence follows by Assumption (iii) in the proposition. It then follows by the Lindeberg-Feller theorem and consistency of the denominator to $\sigma^2_*$ that, for each $c \in \mathbb{R}$,

\begin{equation}
\mathbb{P}[T_1^* \leq c|\boldsymbol{X},g_J] \overset{p}{\to}\Phi(c),
\end{equation}
as desired.
\end{proof}

\subsubsection{Proof of Proposition \ref{Prop_asympt2}}
\label{Proof_Prop_asympt2}

	\begin{proof}
		That $\frac{N}{\sqrt{v_J}}(\hat{\beta}^*_{SS} -b_J) = O_P(1)$ follows from Assumption (i) in the statement of the proposition and convergence in distribution of the numerator of the shift-share regression estimator rescaled by $\sqrt{v_J}$, which we proved in the previous proposition. Moreover, in light of the previous proposition, to conclude that the test statistic converges in probability, it suffices to show that estimation of the residuals using $\hat{\beta}^*_{SS}$ does not asymptotically affect consistency of the variance estimator to $\sigma^2_*$. Specifically, it suffices to verify that, under the null,
		
		\begin{equation}
			\begin{aligned}
		\frac{1}{v_J}	\sum_{j=1}^J \left(\sum_{i=1}^N [\epsilon_{i,J}+\eta_{i,J} + (\beta_J-b_J)X_{i,J}+(b_J-\hat{\beta}^*_{SS} )(\boldsymbol{s}_i'g^*_{J})]\boldsymbol{s}_{i,j,J}\right)^2 g_{j,J}^{*2} = \\ \frac{1}{v_J}	\sum_{j=1}^J \left(\sum_{i=1}^N [\epsilon_{i,J}+\eta_{i,J} + (\beta_J-b_J)X_{i,J}]\boldsymbol{s}_{i,j,J}\right)^2 g_{j,J}^{*2} + o_p(1).
			\end{aligned}
		\end{equation}
	In particular, it is sufficient that
	
	\begin{equation}
		\begin{aligned}
						\frac{1}{v_J}	\sum_{j=1}^J \Bigg(\sum_{i=1}^N \sum_{l=1}^N \Big[(b_J-\hat{\beta}^*_{SS} )s_{i,j,J}s_{l,j,J}[(\boldsymbol{s}_l'g^*_{J})(\epsilon_{i,J}+\eta_{i,J} + (\beta_J-b_J)X_{i,J}) + \\ (\boldsymbol{s}_i'g^*_{J})(\epsilon_{l,J}+\eta_{l,J}) + (\beta_J-b_J)X_{i,J}] + (b_J-\hat{\beta}^*_{SS} )^2 (\boldsymbol{s}_i'g^*_{J}) (\boldsymbol{s}_l'g^*_{J})s_{i,j,J}s_{l,j,J} \Big]\Bigg) g_{j,J}^{*2} = o_p(1),
		\end{aligned} 
	\end{equation}
	which is ensured by assumptions (ii) and (iii).
	\end{proof}

\subsection{Examples of simulation distributions}

\subsubsection{Verification of conditions of Proposition \ref{Prop_asympt}}

\label{Ver_prop_asympt}

We verify the conditions of \ref{Prop_asympt} in three examples.

\paragraph{Nonparametric bootstrap} In this case, $g^*_{j,J} \overset{iid}{\sim}\hat{F}_g$, where $\hat{F}_g$ is the empirical distribution of recentered shocks, i.e. $\hat{F}_g(c)=\frac{1}{J}\sum_{j=1}^J \mathbf{1}_{\{g_{j,J} - \bar{g}_J \leq c\}}$ and $\bar{g}_J=\frac{1}{J}\sum_{j=1}^J g_{j,J}$. In this case, $\mathbb{E}_*[g_{j,J}^*] = 0$, which ensures condition (i). As for the second condition, since $\mathbb{E}_* |g^*_{j,J}|^2 = \frac{1}{J}\sum_{j=1}^J |g_{j,J} - \bar{g}_J|^2$, requirement (ii) in the Proposition will be satisfied if $\frac{1}{J}\sum_{j=1}^J |g_{j,J} - \bar{g}_J|^2$ converges in probability to a positive nonrandom limit and $\sum_{j=1}^J\frac{1}{v_J}\Big( \sum_{i=1}^N [\epsilon_{i,J}+\eta_{i,J}+ (\beta_J-b_J)X_{i,J}]\boldsymbol{s}_{i,j,J}\Big)^2$ converges in probability to a positive nonrandom limit. Finaly, requirement (iii) is satisfied if $\mathbb{E}_* |g^*_{j,J}|^4 = \frac{1}{J}\sum_{j=1}^J |g_{j,J} - \bar{g}_J|^4$ converges in probability and  $\sum_{j=1}^J\frac{1}{{v}_J^2}\left( \sum_{i=1}^N [\epsilon_{i,J}+\eta_{i,J}+ (\beta_J-b_J)X_{i,J}]\boldsymbol{s}_{i,j,J}\right)^4$ converges in probability to zero.

\paragraph{Normal distribution} Suppose $g^*_J \sim N(0, \mathbb{I}_{J\times J})$, independently from $\boldsymbol{X}, g_J$. In this case, $\mathbb{E}_*[g_{j,J}^*] = 0$, for $j=1,\ldots, J$, which ensures condition (i). As for the second condition in the Proposition, $\sum_{j=1}^J\frac{1}{{v}_J}\Big( \sum_{i=1}^N [\epsilon_{i,J}+\eta_{i,J}+(\beta_J-b_J)X_{i,J}]\boldsymbol{s}_{i,j,J}\Big)^2\mathbb{E}_*g_{j,J}^{*2}= \sum_{j=1}^J\frac{1}{{v}_J}\Big( \sum_{i=1}^N [\epsilon_{i,J}+\eta_{i,J}+ (\beta_J-b_J)X_{i,J}]\boldsymbol{s}_{i,j,J}\Big)^2$, which converges in probability if the latter term converges. As discussed in Appendix A of AKM, convergence in probability of $\sum_{j=1}^J\frac{1}{{v}_J}\Big( \sum_{i=1}^N [\epsilon_{i,J}+\eta_{i,J} + (\beta_J-b_J)X_{i,J}]\boldsymbol{s}_{i,j,J}\Big)^2$ to a positive limit requires the existence of at least one ``non-negligible'' shock in most units, where by non-negligible shock in a unit we mean its exposure weight is bounded away from zero. In an ``extreme'' case, where $N=J$ and each unit is affected by a single distinct shock with unit exposure, this term simplifies to $\frac{1}{N}\sum_{i=1}^N[\epsilon_{i,J}+\eta_{i,J}+(\beta_J-b_J)X_{i,J}]^2$, which is expected to converge to a positive limit under mild conditions. Finally, the third condition simplifies to $\sum_{j=1}^J\frac{1}{{v}_J^2}\left( \sum_{i=1}^N [\epsilon_{i,J}+\eta_{i,J} + (\beta_J-b_J)X_{i,J}]\boldsymbol{s}_{i,j,J}\right)^4\mathbb{E}_*g_{j,J}^{*4} = 3 \sum_{j=1}^J\frac{1}{{v}_J^2}\left( \sum_{i=1}^N [\epsilon_{i,J}+\eta_{i,J}+(\beta_J-b_J)X_{i,J}]\boldsymbol{s}_{i,j,J}\right)^4$, which converges to zero if the latter term converges. In the single-shock-exposure setting, this term simplifies to $\frac{1}{N^2}\sum_{i=1}^N [\epsilon_{i,J}+\eta_{i,J} + (\beta_J-b_J)X_{i,J}]^4$, which converges in probability to zero under mild conditions.

\paragraph{Sign changes}
	In this case, $g^*_J = \pi^* \odot g_J$, where $\odot$ denotes entry-by-entry multiplication, and $\pi^* \sim \operatorname{Uniform}(\{-1,1\}^J)$, independently from $(g_J, \boldsymbol{X})$. By construction, $\mathbb{E}_*[g_{j,J}^*] = 0$. As for the second condition, $\sum_{j=1}^J\frac{1}{{v}_J}\left( \sum_{i=1}^N [\epsilon_{i,J}+\eta_{i,J} + (\beta_J-b_J)X_{i,J}]\boldsymbol{s}_{i,j,J}\right)^2\mathbb{E}_*g_{j,J}^{*2} = \sum_{j=1}^J\frac{1}{{v}_J}\left( \sum_{i=1}^N [\epsilon_{i,J}+\eta_{i,J}+(\beta_J-b_J)X_{i,J}]\boldsymbol{s}_{i,j,J}\right)^2 g_{j,J}^2$, which converges in probability to a positive constant if: (a) $\sum_{j=1}^J\frac{1}{{v}_J}\left( \sum_{i=1}^N [\epsilon_{i,J}+\eta_{i,J}+(\beta_J-b_J)X_{i,J}]\boldsymbol{s}_{i,j,J}\right)^2 \mathbb{E}[g_{j,J}^2]$ converges to a positive constant; and (b) $\mathbb{E}\left|\sum_{j=1}^J\frac{1}{{v}_J}\left( \sum_{i=1}^N [\epsilon_{i,J}+\eta_{i,J}+(\beta_J-b_J)X_{i,J}]\boldsymbol{s}_{i,j,J}\right)^2 (g_{j,J}^2-\mathbb{E}[g_{j,j}])^2\right|^2$ converges to zero. A condition like (a) is required for existing inference methods in shift-share designs to work (see the discussion surounding Assumption A.1. in AKM). Condition (b) is satisfied if shocks $g_{j,J}$ are independent, $\mathbb{E}|g_{j,J}|^4$ is uniformly bounded and $\sum_{j=1}^J\frac{1}{{v}_J^2}\left( \sum_{i=1}^N [\epsilon_{i,J}+\eta_{i,J}+(\beta_J-b_J)X_{i,J}]\boldsymbol{s}_{i,j,J}\right)^4$ converges in probability to zero. Finally, condition (iii) in the Proposition is satisfied if $\mathbb{E}|g_{j,J}|^4$ is uniformly bounded and $\sum_{j=1}^J\frac{1}{{v}_J^2}\Big( \sum_{i=1}^N [\epsilon_{i,J}+\eta_{i,J}+(\beta_J-b_J)X_{i,J}]\boldsymbol{s}_{i,j,J}\Big)^4$ converges to zero.

\subsubsection{Verification of conditions of Proposition \ref{Prop_asympt2}}

\label{Ver_prop_asympt2}

We now discuss Assumptions (i)-(iii) of Proposition \ref{Prop_asympt2} in the context of the three examples in the previous section.

\paragraph{Verification of condition (i) of Proposition \ref{Prop_asympt2}}
	When the distribution of simulated shocks is standard normal, $\frac{1}{N} \sum_{i=1}^N  \mathbb{E}[(\boldsymbol{s}_{i,J}'g_J^*)^2] = \frac{\sum_{i=1}^N \sum_{j=1}^J s_{i,j,J}^2}{N}$, which we require to converge to a positive constant. Such condition is analogous to Assumption A1.(ii) in AKM. Moreover, we note that, in the single-exposure case, this quantity is exactly equal to one. To conclude that the denominator of the simulated shift-share regression estimator converges in probability to a positive constant, it is sufficient to require that $\mathbb{V}\left[\frac{1}{N} \sum_{i=1}^N (\boldsymbol{s}_{i,J}'g_J^*)^2 \right] = o(1)$. Observe that, in the single-exposure case, this variance is given by $\frac{ 3N + N(N-1)}{N^2} - 1$, which converges to zero as $N \to \infty$. Similar arguments establish convergence of the denominator of the shift-share regression estimator in bootstrap and sign changes examples.

\paragraph{Verification of condition (ii-iii) of Proposition \ref{Prop_asympt2}}

 Assumptions (ii-iii) implicitly restrict moments of the simulated shocks and the relation between exposure weights and the rate of growth of $N$. Indeed, in the single-exposure case, requirement (ii) subsumes to
	
	\begin{equation}
		\frac{2}{N^{3/2}} \sum_{i=1}^N g_{j,J}^{*3}(\epsilon_{j,J} + \eta_{j,J} + (\beta_J-b_J)X_{i,J}) = o_p(1),
	\end{equation}
	which is expected to hold under mild conditions in our three main examples. Similarly, in the single exposure case, condition (iv) subsumes to,
	
	\begin{equation}
		\frac{1}{N^{2}} \sum_{i=1}^N g_{j,J}^{*4}= o_p(1),
	\end{equation}
	which is also expected to hold.

\end{document}